# Nearfield Vortex Dynamics of Supercell Bloch Modes

Xiaona Ye[1†], Guangfeng Wang[1†], Xiaoyang Duan[2]*, Ziwei Wang[1], Zengya Li[1], Tongtong Jia[3], Tingxin Li[3], Luqi Yuan[1], Bo Wang[1]*, Xianfeng Chen[1,4,5]*

[1]State Key Laboratory of Advanced Optical Communication Systems and Networks, School of Physics and Astronomy, Shanghai Jiao Tong University, Shanghai, 200240, China.
[2]Key Laboratory of Advanced Optoelectronic Quantum Architecture and Measurement (Ministry of Education), Beijing Key Laboratory of Nanophotonics and Ultrafine Optoelectronic Systems, School of Physics, Beijing Institute of Technology, Beijing 100081, China.
[3]Key Laboratory of Artificial Structures and Quantum Control (Ministry of Education), School of Physics and Astronomy, Shanghai Jiao Tong University, Shanghai 200240, China.
[4]Shanghai Research Center for Quantum Sciences, Shanghai, 201315, China.
[5]Collaborative Innovation Center of Light Manipulations and Applications, Shandong Normal University, Jinan, 250358, China.

†These authors contributed equally to this work.
*Corresponding author. Email: xduan@bit.edu.cn (X.D.); wangbo89@sjtu.edu.cn (B.W.); xfchen@sjtu.edu.cn (X.C.)

## Abstract

Densely arranged optical vortices are natural solutions of high-symmetry Bloch modes in photonic crystals. However, strict symmetry constraints limit the potential spatial configurations of nearfield vortices, restricting the control over light-matter interaction. Here, we demonstrate a nearfield vortex dynamic within a supercell photonic crystal. By introducing paired rotations of triangular structures, we achieve high-quality-factor Bloch mode transition from evanescent valley modes, to quasi-bound states in the continuum, frustrated modes, and quasi-valleys. Each stage exhibits distinct nearfield vortex distributions, nonlinear overlap properties, and quality factors, revealing diverse physical behaviors for tailoring light-matter interaction. Notably, the asymmetric vortex configuration of frustrated modes enhances second harmonic generation, driven by an optimized nonlinear overlap factor. Our paired-rotation strategy offers a versatile design framework for creating supercell photonic crystals with unique nearfield vortex properties, presenting promising applications in lasing, nonlinear optics and optical forces.

**Introduction.** The strong light-matter interaction achieved from high-quality ($Q$) factor Bloch modes of photonic crystal slabs (PCSs) can boost many optical effects within a subwavelength thickness, with many applications in laser [1, 2], single photon generation [3], and high-order harmonic generation [4–6]. The exploiting of group theory and parameter perturbation method has provided a paradigm for designing novel PCSs [7–9]. Multiple design approaches, including BIC merging [10], Brillion-zone folding [11], twisted or Moiré PCSs [12–15], and engineered rotational symmetry [16,17], have been proposed to enrich the parameter space for designing PCSs with different functionalities. In particular, the rotation of geometry structures has shown as a powerful means to flexibly tune the optical properties of PCSs. For instance, the slightly rotating of structures can break $C_2$ symmetry, inducing quasi-bound states in the continuum (BICs) [18,19]. For honeycomb PCSs, the rotation of $C_3$ structure will open the Dirac point to form topological valley pairs [20,21]. Moreover, spatially-varying rotation of $C_2$ structures gives rise to real-space geometric phases for customized laser radiation [22–24].

A particular case of utilizing PCSs for enhanced light-matter interaction is to generate second harmonic (SH), which is usually a very weak effect that requires a sufficiently long accumulation distance to obtain a fair conversion efficiency. For PCSs and other optical structures of a similar thickness [25–30], the SH accumulation is achieved by enhanced $Q$ factor. Notably, the asymmetric nature of SH generation also requires a strong nonlinear field overlapping ($\beta$) between the fundamental mode and SH mode [31]. However, optical structures with symmetry-guaranteed high-$Q$ modes usually result in very small $\beta$ [32]. For instance, the extremely high-$Q$ factor of quasi-BICs (qBICs) emerges from structures with a $C_{2n}$ symmetry, while a considerably large $\beta$ arises from those with a $C_{2n+1}$ symmetry [33–35]. This $Q$–$\beta$ contradiction has become a main barrier that hinders us to find an optimized solution for SH generation in PCSs and metasurfaces. Therefore, seeking a balance between spatial symmetry breaking and high-$Q$ effect are critical for optimized SH generation in resonate structures, which requires novel design strategies beyond conventional photonic devices.

In this work, we designed a $C_3$ supercell PCS and observed unique nearfield vortex dynamics. The supercell PCS features three triangular meta-atoms arranged in a hexagonal lattice, with a pair of opposite rotations of the bottom triangles tuning the in-plane rotational symmetry and valley mode coupling. As the bottom triangles rotate, three free vortices in the supercell move towards the top triangle. During this process, distinct Bloch modes are revealed from different vortex distributions. Four typical stages are valley modes, qBICs, frustrated modes (FMs), and quasi-Valleys (qValleys). These Bloch modes are characterized by different nearfield vortex formation, quality factor, field asymmetry, and even energy flow, providing a very rich realm to modulate light-matter interactions by simply rotating the unitcell structures. In particular, there is a critical point where four vortices are merged, forming degenerate "two-down-one-up" and "two-up-one-down" frustration-like states (Fig. 1b), with "up" or "down" being the handedness of the phase vortex. This pseudospin feature is an optical analogy to geometric frustration, which exhibits similar asymmetric spin (magnetic) configuration [36]. The FM is associated with a maximum $\beta$ from its unique asymmetric vortex formation. Meanwhile, the radiative $Q$ factor reaches a minimum, corresponding to a positive topological charge in the parameter space. The vortex distribution in the supercell has guaranteed a considerably large $Q\beta$, making the FM as an optimized solution for SH generation.

**Bloch mode dynamics in the supercell PCS.** We start the theoretical analysis by examining the transverse electric modes of a PCS consisting of a hexagonal lattice of triangular meta-atoms (Fig. 2a and b). Initially, the triangles are uniformly rotated from the $y$ axis by an angle $\delta$, which determines the PCS's symmetry and band degeneracy. Specifically, for $\delta = (2m+1)\pi/6$ ($m \in \mathbb{Z}$), the point group at the corners ($\pm K$) of the Brillouin zone exhibits $C_{3v}$ symmetry, corresponding to double-degenerate Dirac modes (Fig. 2e, the black bands). Otherwise, the symmetry reduces to $C_3$, and the two-fold Dirac degeneracy is lifted to form a valley pair ($\psi_1$ and $\psi_2$ at $\pm K$ in Fig. 2e) [37–39].

Based on this $\delta$-dependent valley PCS, we introduce a supercell structure composed of three neighboring unitcells (Fig. 2c). In the supercell, the rotation angle of the top triangle is set as $\delta_T$, while the bottom triangles are characterized by a pair of opposite angles, $\pm\delta_B$. This design strategy is inspired by a natural non-centrosymmetric compound HoAgGe, where the structural distortion is similarly characterized by opposite rotations of two types of triangles to induce frustration [40]. Two straightforward outcomes from the supercell PCS are (I) the reduction of the Brillion zone (Fig. 2d) and (II) band structure folding from the K points to Γ points with valley mode coupling (Fig. 2e). Specifically, in the supercell structure, the valley pairs $\psi_1$ and $\psi_2$ become two pairs of modes, $\Gamma_{1,2}$ and $\Gamma_{3,4}$. Their eigen frequencies are determined by $\delta_T$ and $\delta_B$, which induces distinct valley mode coupling (Fig. 3c). The double degeneracy will always be remained due to the supercell $C_3$ symmetry.

As an example, we evaluated the $Q$-factor evolution of the $\Gamma_{1,2}$ mode via simultaneously rotating $\delta_T$ and $\delta_B$. The radiative loss ($1/Q$) of $\Gamma_{1,2}$ mode is shown in Fig. 3(a), with the minus and plus signs indicating two extreme points. In fact, for $\delta_T = \delta_B = 0$, the $\Gamma_{1,2}$ are K-point valley modes, which have an infinite $Q$ factor as they are residing in the evanescent region. Any slight change of ($\delta_T$, $\delta_B$) around the valley mode will induce a sudden transition from valley modes to paired qBICs, resulting in a very high but finite $Q$ factor. Therefore, the valley mode exhibits a negative topological charge $q = -1$, which is defined in the parameter space by $q = \oint \nabla(1/Q) \cdot d\vec{l}$. Here, $\nabla = (\partial/\partial\delta_T, \partial/\partial\delta_B)$, and $\vec{l}$ is a differential path on the ($\delta_T$, $\delta_B$) surface [the arrows in Fig. 3b are $\nabla(1/Q)$]. The qBIC region is characterized by a quadratically decayed $Q$ factor along with the rotational parameters, that is, $Q \propto (\delta_B^2 + \delta_T^2)^{-1}$. This qBIC region is approximately sketched by the circle in Fig. 3b. In contrast, a largest radiative loss occurs around $(\delta_T, \delta_B) = (0, 40°)$, corresponding to a positive topological charge $q = 1$. This minimal $Q$ factor is equivalent to a maximum of radiation loss, as $|\kappa| \propto 1/\sqrt{Q}$ [41]. This mode is also referred to as a FM — a unique vortex-merged Bloch state that will be explained later. Notably, the symmetry constrain of the supercell requires that the topological charge $q$ can only occur for $\delta_T = m\pi/3$ (section S7). Therefore, there is only one pair of opposite topological charges in the $(\delta_T, \delta_B)$ space (Fig. 3b).

A clear mode evolution can be visualized from the supercell vortex dynamic. Figure 3d shows several typical nearfield examples of $\Gamma_1$ with $H_z$ phase vortices. The vortices of valley modes are periodically arranged in the PCS with an equal distance. One group of vortices reside at the centers of the triangles (Fig. 3d left panel, the circled arrows), while the other group of opposite vortices are located at three corners of the supercell boundaries (Fig.3d, the white dots). We define the vortices at the centers of the triangles as pseudospin vectors, $\mathbf{S} = \pm\mathbf{z}$, with $\mathbf{z}$ being the direction of $\mathbf{S}$, corresponding to a clockwise or anticlockwise phase vortex, respectively. The vortices at the corners of the supercells are referred to as free vortices, as they will move in the nearfield once we change $\delta_B$. Pseudospin vortices always remain at the center of the triangles, and they can be utilized to characterize valley index [23]. In contrast, as $\delta_B$ changes, free vortices move towards the top triangle with different displacements from their original locations, resulting in distinct Bloch states (section S8).

In term of vortex distribution, there are three other typical states besides valley modes. The qBICs are Bloch modes with their free vortices slightly deviated from that of the valley modes (Fig. 3d, second panel). The qValley mode occur for $\delta_B = 60°$, which exhibits the same vortex distribution with valley mode, but with a different spatial intensity periodicity (Fig. 3d, fourth panel). There is a critical point of $\delta_B$, at which the free vortices are merged with the upper pseudospin, resulting in a FM with "two-down-one-up" or "two-up-one-down" supercell vortex formation, akin to a magnetic frustration. Here, "up" or "down" is defined by the sign of $\mathbf{S} = \pm\mathbf{z}$, respectively. The critical point $\delta_B \in (0°, 60°)$ is determined by the geometric parameters of the PCS. Notably, because of this unique vortex distribution in the supercell, FM is associated with the highest radiative loss ($q = 1$) and a strong spatial asymmetry in the electric field distribution (the lower panel of Fig. 3d). The symmetry constrain of the supercell also requires that the FM can only occur for $\delta_T = m\pi/3$ (section S7).

A potential application of the FM is SH enhancement from its spatial asymmetry. Experimentally, we achieved a $10^6$ enhancement from a FM compared to a Si film, which does not generate SH from bulk material. Figure 3(e) shows the simulated results of SH enhancement on the $\Gamma_{1,2}$ band from on-resonant pumping with a circularly polarized plane wave. Notably, the highest SH enhancement occurs around FM, rather than qBICs (although they exhibit the highest finite $Q$ factors). Based on temporal coupled-mode theory, we find that FM mode has a maximum $\beta$ associated with its unique vortex distribution (section S9). Moreover, Fig. 3(b) and (e) also show a strong SH generation in a wide $\delta_B$ range around the FM, indicating a robust performance against the rotation of triangles. An important criterial is that the SH increases as $\delta_B$ changes from 0 to 20°, then the SH generation becomes stable.

**Experimental characterization of supercell PCS enhanced SHG.** Experimentally, we fabricated a series of PCSs with varying geometric parameters using a silicon-on-insulator device (section S1). The thickness of the silicon film is $H = 220$ nm. The lattice constant of the PCS is $a = 315$ nm, and the circumcircle radius $R$ of the triangles ranges from 141 nm to 153 nm, with several $\delta_B$ selectively chosen from 0 to 20° (section S5). Typical scanning electron microscope images are shown in Fig. 4a for the case of $\delta_B = 20°$. We used a cross-polarized reflectance spectral characterization to measure the spectra of different PCSs [Fig.4(b)], which are in good agreement with the simulations. Moreover, the wavelength of the resonant mode and $Q$-factors are systematically characterized by measuring different PCSs with varying $R$ and $\delta_B$ [Fig.4(c) and (d)]. Particularly, the change of $\delta_B$ from 0° to 20° will reduce the $Q$ factor from ~2000 to ~200, indicating a mode transition from valley, via qBICs, towards FM.

We experimentally characterized the SH of the supercell PCSs using a pulsed laser (~8 ps) with a central wavelength of 1064 nm and a repetition rate of 20 MHz. The pump light is circularly polarized with a focused spot size of ~2 μm, impinging onto the PCSs with a normal incident angle (section S3). Typical spectra for the pump laser and the SH are presented in Fig. 5a. Figure 5b illustrates the measured SH power as a function of the pumping power, showing a quadratic relation between the pump and signal. The measured SH in the real space is shown in Fig. 5(c) by scanning the location of pumping beam. The intensity profile indicates an enhanced field of the Bloch mode in the center of PCS.

Figure 5d shows the experimental results of the SH power in supercell PCSs with different $R$ and $\delta_B$, pumped by a laser power of 14.3 mW. Ideally, the valley modes do not generate SH from the Γ-point pump. However, the nanoscale roughness in the fabricated meta-atoms naturally induces spatially variant perturbations that create a momentum channel for SH. The results presented in Fig. 5e span different Bloch modes for the wavelength of interest, including qBICs and FMs, via different $\delta_B$. Among them, a maximum SH power enhancement factor of ~340 is achieved between a qBIC ($R = 149$ nm, $\delta_B = 5°$) and a FM ($R = 147$ nm, $\delta_B = 20°$), with a maximum power of ~891 pW. In general, we observe about 2 orders of magnitude SH enhancement as $\delta_B$ increases from 0 to 20° (Fig. 5e). These results demonstrate the robust performance of FM in SH generation against parameter changes, in good agreement with the theoretical prediction (section S10). Notably, the highest SH from the FMs compared to that from a bared Si film is larger than $10^6$ (Fig. 5e).

**Discussion**.

In conclusion, we have introduced a supercell PCS featuring paired rotations of triangular meta-atoms, inspired by the structural distortion of natural non-centrosymmetric compounds like HoAgGe, where two types of triangular units rotate oppositely. This paired rotation induces a continuous nearfield vortex dynamic for paired Bloch modes, which maintain a double mode degeneracy at the Γ point due to the supercell's $C_3$ symmetry. This approach enables flexible tuning of Bloch mode properties, including quality factor, momentum (Brillion-zone folding), spatial symmetry, and nonlinear overlap factor. Different stages of these modes are distinctly defined by unique nearfield vortex distributions. We show an optical analogy of FM within the PCS, presented as degenerate "two-down-one-up" and "two-up-one-down" pseudospin vortex textures. These FMs serve as an optimal condition for SH generation by maximizing the $Q\beta$ factor, a mechanism that contrasts with SH generation from qBICs. Our rotational design strategy can be broadly applied, including the creation of larger supercells or the extension from in-plane rotation to out-of-plane tilting, offering a versatile framework for designing supercell photonic crystals with tailored properties and promising applications in nonlinear optics, optical forces, and laser effect.

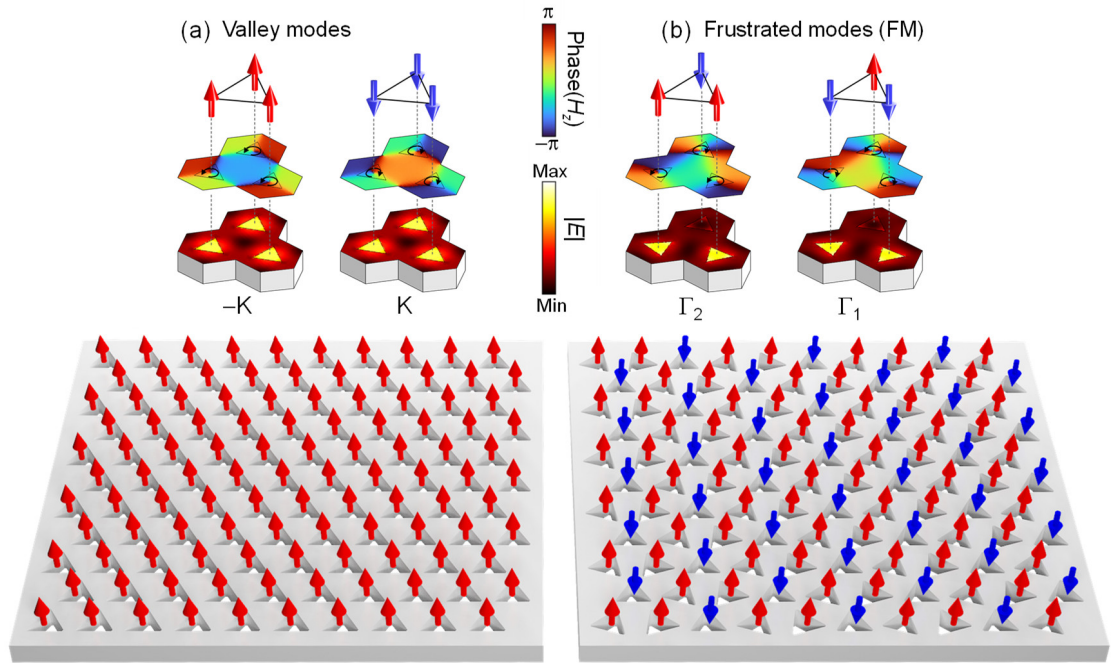

**Fig. 1 Concept illustration of frustrated photonic Bloch modes**. (a) The ±K-valley modes of a conventional PCS consisting of a hexagonal array of triangular meta-atoms. (b) Degenerate frustrated modes in a supercell PCS. The arrows represent pseudospin vectors defined by the local phase vortices at the centers of triangular meta-atoms.

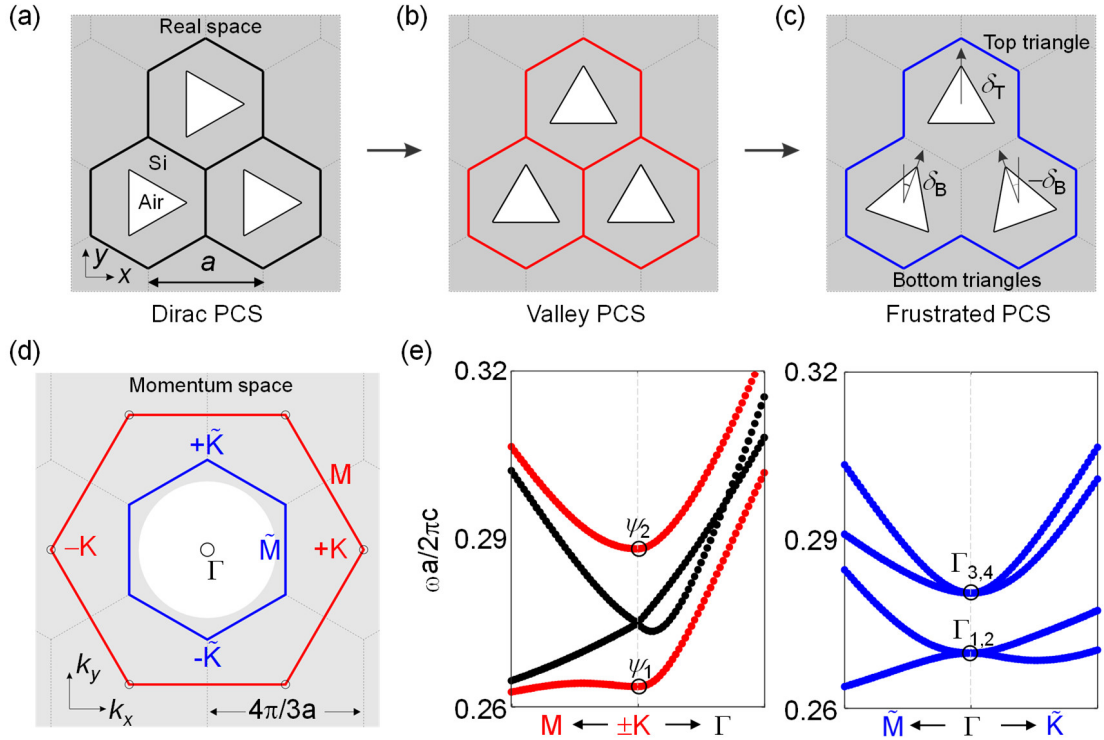

**Fig. 2 Supercell PCS and corresponding band structures. (a), (b)** Structures of conventional PCSs consisting of triangular meta-atoms with uniform rotation angles of $\delta = \pi/6$ (Dirac) and $\delta = 0$ (Valley), respectively. **(c)** Designed structure of a frustrated supercell PCS. The angle of the top triangular meta-atom is $\delta_T = 0$, and the angles of the bottom triangles are $\pm\delta_B$. **(d)** Brillouin zones for the valley PCS (red hexagon) and the supercell PCS (blue hexagon). The shaded area represents the waveguide region. **(e)** TE band structure of the Dirac/valley PCSs (red and black dots) around $\pm K$ points and the supercell PCSs (blue dots) around $\Gamma$ point.

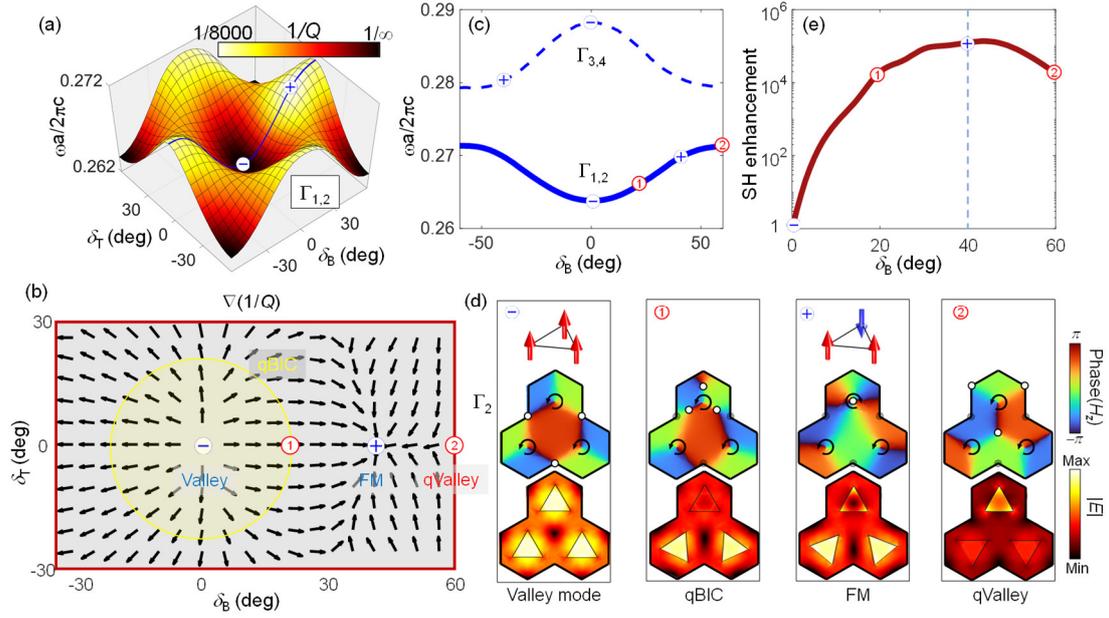

**Fig. 3 Evolution of Bloch modes and SH enhancement in parameter space. (a)** The dispersion surface of $\Gamma_{1,2}$ modes and the corresponding radiation loss ($1/Q$) mapped in the parameter space ($\delta_T, \delta_B$). The "−" and "+" signs are locations with a minimal and maximal $1/Q$. **(b)** The vector map of $\nabla(1/Q)$ in the rotation parameter space. The yellow circular region approximately indicates the area for qBICs. **(c)** The eigen frequency evolution of two pairs of $\Gamma$-point modes as a function of $\delta_B$, for $\delta_T = 0°$. The dashed curve represents the higher pair of modes, $\Gamma_{3,4}$. **(d)** Selected pseudospin textures **S** (top panel), nearfield vortex distributions (middle panel), and electric field distributions (bottom panel) for four $\Gamma_2$ modes in corresponding to the marks in (c). The transparent black dots are the original locations of the free vortices starting from $\delta_B = 0$. **(e)** Simulated SH enhancement as a function of $\delta_B$, which is calculated by exciting the $\Gamma_{1,2}$ modes with a circularly polarized plane wave. These vortices are shown by decomposing the paired Bloch modes into a pair of opposite circularly-polarized modes.

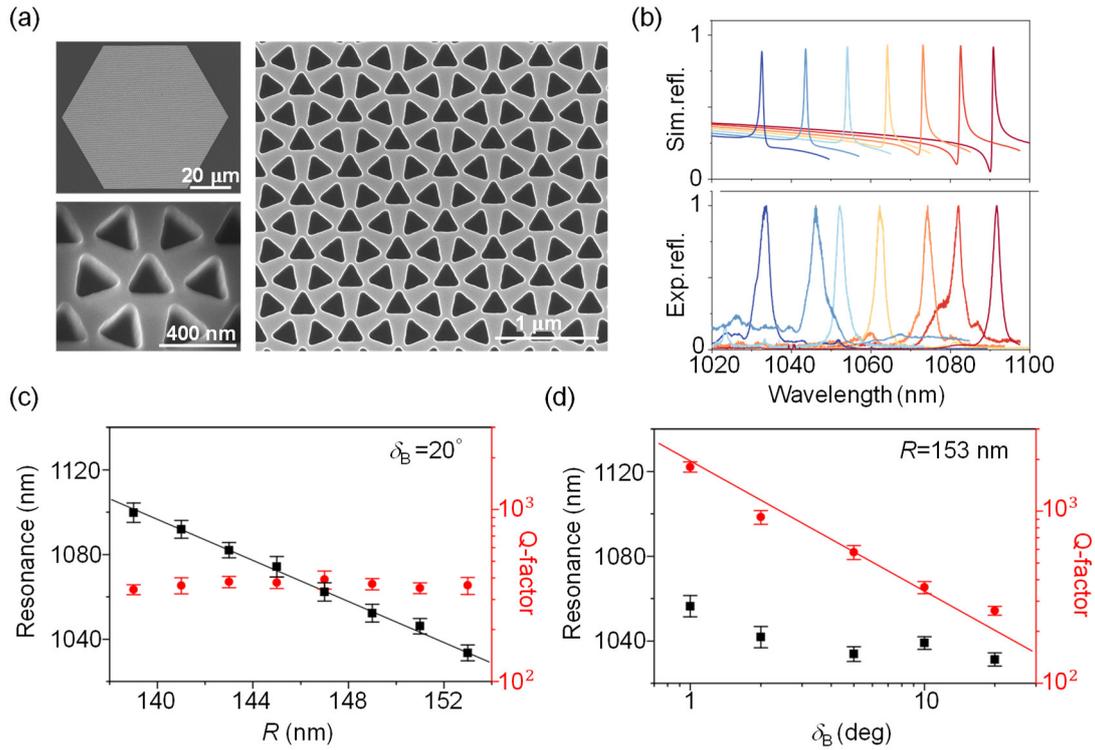

**Fig. 4 Spectral characterization of supercell PCSs.** (**a**) SEM images of a typical supercell PCS with $\delta_B = 20°$. (**b**) Simulated (upper panel) and measured (lower panel) reflectance spectra of the Si PCSs with various radii $R$. The $R$ decreases from 153 nm to 141 nm in a step of 2 nm from the leftist to the rightest spectrum. The lattice constant is $a = 315$ nm and $\delta_B = 20°$. (**c**) Measured resonance wavelengths and $Q$ factors for supercell PCSs with different $R$. (**d**) Measured resonance wavelengths and $Q$ factors as a function of $\delta_B$.

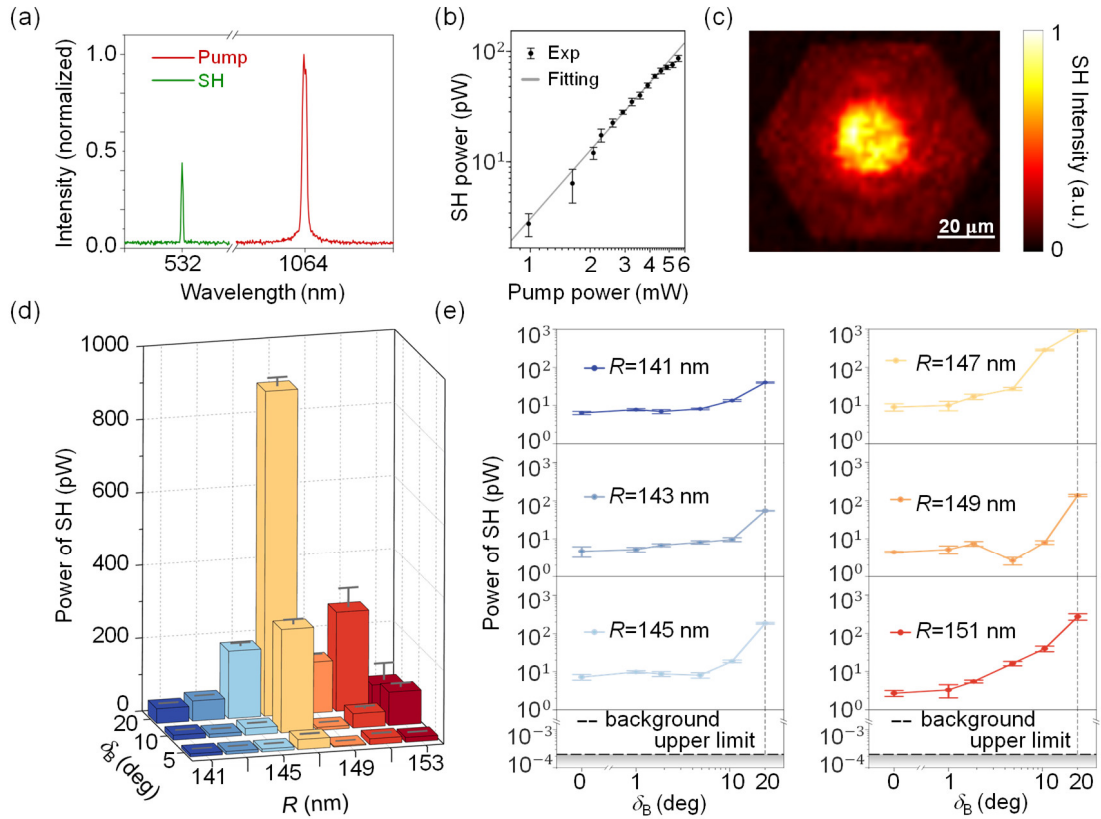

**Fig. 5. Experimental characterization of the SH enhancement from supercell PCSs.** (**a**) Typical spectrum of fundamental wave (red) and SH generation (green). (**b**) Power dependence of SH on a logarithmic scale, showing quadratic power scaling. The black dots represent the measured data, and the line is fitted with a second-order power function. (**c**) Measured SH by scanning the pumping laser over the surface of a supercell PCS for $R$ = 147 nm and $\delta_B$ = 20°. This intensity distribution indicates the spatial profile of the Bloch modes from the PCS. (**d**) Measured SH power as a function of $R$ and $\delta_B$ from a circularly polarized pump beam with a power of 14.3 mW. (**e**) Selected experimental SH generation as functions of $\delta_B$. The background upper limit (horizontal dashed lines) is the SH signal measured from Si film, which is ~6 orders of magnitudes smaller than the maximum SH enhancement from the PCSs.

**Acknowledgements**

This work is supported by National Key R&D Program of China (2022YFA1205101), National Science Foundation of China (12274296, 12192252), Shanghai International Cooperation Program for Science and Technology (22520714300), Shanghai Jiao Tong University 2030 Initiative, Beijing Municipal Natural Science Foundation (Z240005). B.W. is sponsored by the Yangyang Development Fund. The samples were fabricated in the center for advanced electronics materials and devices (AEMD), Shanghai Jiao Tong University.

# Supplementary Materials for

# Nearfield Vortex Dynamics of Supercell Bloch Modes

Xiaona Ye[1†], Guangfeng Wang[1†], Xiaoyang Duan[2]*, Ziwei Wang[1], Zengya Li[1], Tongtong Jia[3], Tingxin Li[3], Luqi Yuan[1], Bo Wang[1]*, Xianfeng Chen[1,4,5]*

[1]State Key Laboratory of Advanced Optical Communication Systems and Networks, School of Physics and Astronomy, Shanghai Jiao Tong University, Shanghai, 200240, China.

[2]Key Laboratory of Advanced Optoelectronic Quantum Architecture and Measurement (Ministry of Education), Beijing Key Laboratory of Nanophotonics and Ultrafine Optoelectronic Systems, School of Physics, Beijing Institute of Technology, Beijing 100081, China.

[3]Key Laboratory of Artificial Structures and Quantum Control (Ministry of Education), School of Physics and Astronomy, Shanghai Jiao Tong University, Shanghai 200240, China.

[4]Shanghai Research Center for Quantum Sciences, Shanghai, 201315, China.

[5]Collaborative Innovation Center of Light Manipulations and Applications, Shandong Normal University, Jinan, 250358, China.

†These authors contributed equally to this work.

*Corresponding author. Email: xduan@bit.edu.cn (X.D.); wangbo89@sjtu.edu.cn (B.W.); xfchen@sjtu.edu.cn (X.C.)

**Numerical simulation**

All electromagnetic simulations, including PCS band structure, SH, and reflectance spectra, were performed using the finite element method with COMSOL Multiphysics. We utilized the frequency domain to calculate the SH generation from silicon PCS. A polarized plane wave at the fundamental frequency ($\omega$) is normally incident onto the supercell structure (with Floquet periodic boundary condition around the structure, and scattering boundaries at top and bottom); the resulted nearfield is $\mathbf{E}(\omega)$. Since bulk silicon is centrosymmetric, there is no SH response within the material. Therefore, we set a thin layer (10 nm) of silicon at the air-silicon interface with an effective $\chi_2$ to generate SH. The source of SH is calculated from $\sim\chi_2\mathbf{E}^2(\omega)$. The power of SH is collected by setting a port at the reflection side of the model, which calculate the radiated power from the PCS at the frequency of $2\omega$.

**Section S1: Sample fabrication**

For the fabrication of supercell photonic crystals, we used a silicon-on-insulator (SOI) wafer, which is consisted of a 725-μm-thick silicon substrate, a 2-μm-thick buried silicon oxide, and a 220-nm-thick silicon. The fabrication process is schematically illustrated in Fig. S1. The wafer was diced into 1 cm × 1 cm pieces, which were cleaned in acetone, isopropanol (IPA), SC-3 cleaner and deionized (DI) water for 20 minutes. Subsequently, a layer of electron beam resist (AR-P 6200.13) was spin-coated on top of the Si film at a speed of 4,000 rpm to achieve a thickness of ~400 nm. The resist was then baked on a hot plate at 150°C for 2 minutes. The patterns were exposed using electron beam lithography (EBL, Vistec EBPG-5200), followed by development in MIBK solution for 75 seconds, and rinsing with IPA for 60 seconds. After lithographic patterning, inductively coupled plasma (ICP) etching (ICP-SR) was performed to etch the Si through the patterned mask, transferring the pattern from the mask to the silicon. Finally, the residual resist was removed using acetone, IPA and DI water, followed by $O_2$ plasma treatment (500 W, 3 minutes), leaving only nanoholes on the Si film.

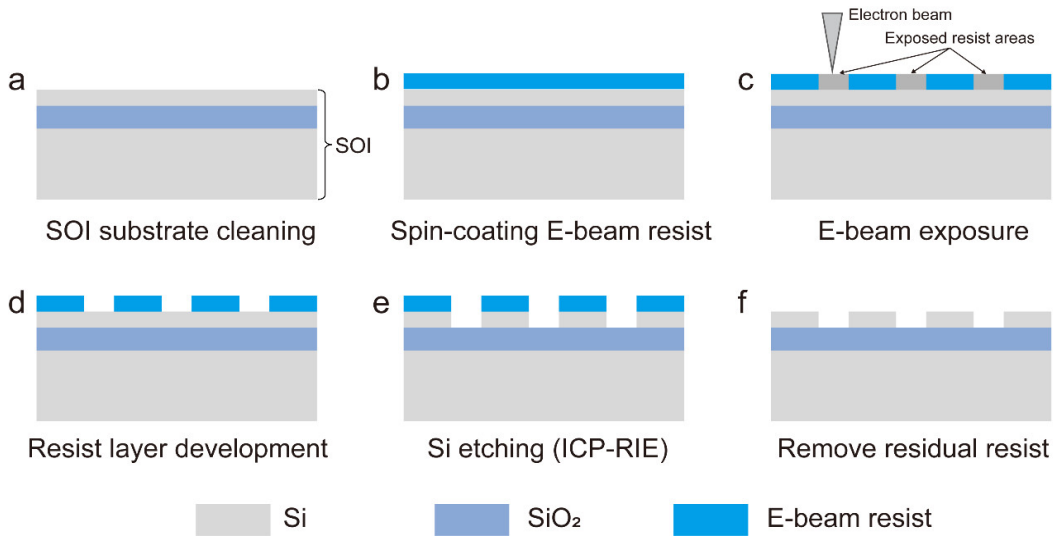

***Fig. S1.*** *Fabrication process of the samples. (a) Wafer cleaning. (b) Spin-coating of e-beam resist. (c) E-beam lithography. (d) Development of the resist layer. (e) Transfer of the pattern to the Si film by dry etching. (f) Removal of residual resist.*

**Supplementary Section S2: Methods for reflectance spectrum**

The reflection measurements were performed using a home-built setup, as shown in Fig. S2. The samples were illuminated by a focused broadband supercontinuum laser (Yangtze Soton Laser Photonics, SC-Pro) via a 20× near-infrared objective with a numerical aperture (NA) of 0.4 (Mitutoyo M Plan APO NIR). The reflected light was collected by the same objective. To filter out background reflection spectra, the illumination beam was set to a specific polarization state (linear polarization), and only the cross-polarized reflected light was collected. Finally, the reflected light was collected by a spectrometer (Zolix, Omni-λ500i) through a multimode fiber. The measured data were fitted using Gaussian or Lorentzian functions to estimate the quality-factor ($Q$-factor).

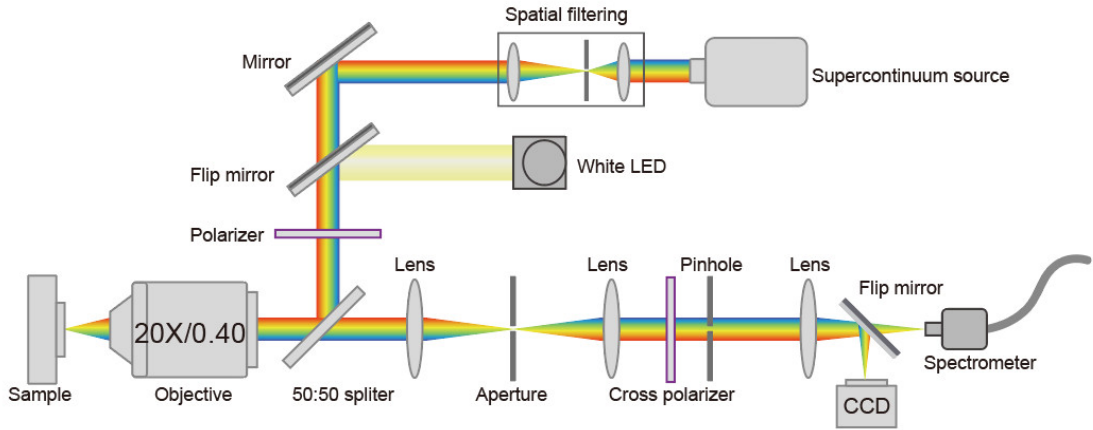

*Fig. S2.* *Schematic of the experimental setup for reflectance spectrum measurement.*

**Supplementary Section S3: Methods for SH measurement**

Optical measurements of second harmonic (SH) generation were performed using a home-built microscopic system, as shown in Fig. S3. The photonic crystals were pumped by a picosecond pulsed laser (Precilasers, FL-PS-1064-8-1.5) with a central wavelength of 1064 nm, a repetition rate of 20 MHz, and a pulse duration of 8 ps. The pump power was adjusted using a polarizer and a halfwave plate. A 50× microscope objective with an NA of 0.42 (Mitutoyo M Plan APO NIR) was used to focus the pump beam. The emitted SH signal and reflected fundamental wave (FW) from the sample were collected using the reflection geometry by the same objective. These signals were spectrally separated using a dichroic mirror and a set of optical filters to remove the fundamental wave. The signal was collected by a visible CCD and a spectrometer (Zolix, Omni-λ500i) for analysis. During the measurement, linear polarizer and quarter-wave plate were used to generate circularly or linearly polarized pump.

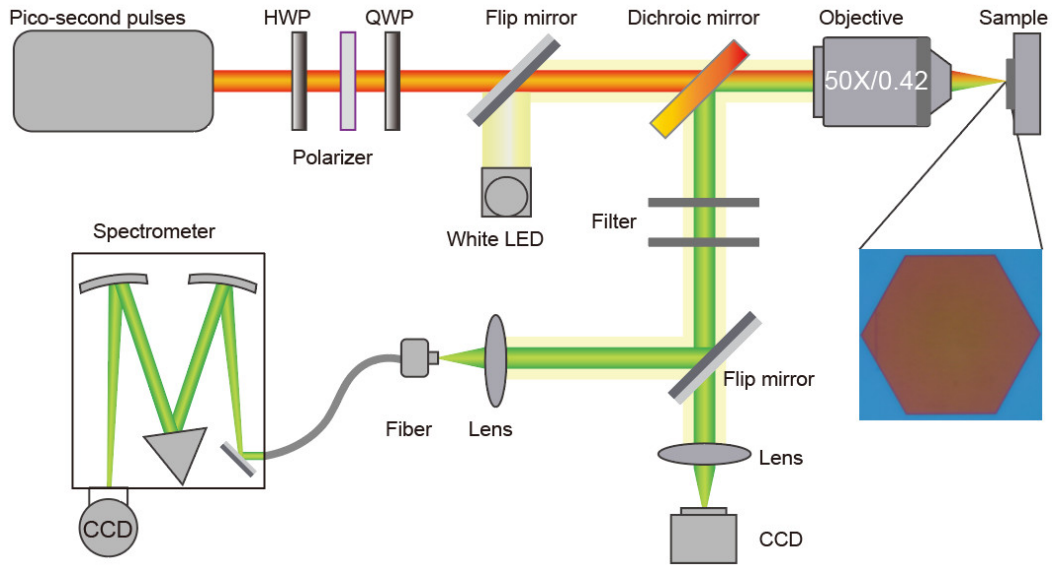

***Fig. S3.*** *The experimental setup for second harmonic generation. HWP, half-wave plate; QWP, quarter -wave plate.*

**Supplementary Section S4: Polarization-resolved SH measurement**

The optical setup for the polarization-resolved measurement is shown in Fig. S4a. The control of input polarization and laser intensity was achieved using a polarizer and a half-wave plate, which are combined to ensure a constant power for different pumping polarization. The analyzer is an electric controlled linear polarizer. During the measurements, the incident polarization and the analyzer were simultaneously rotated over 180°, while their angles were kept in either parallel or perpendicular directions. Figure S4b shows a schematic of the measured supercell PCS. The polarization-resolved SH generations are shown in Fig. S4c, with four cases presented. It can be seen that the PCS is not sensitive to circular polarization, as the perpendicular and parallel circular polarization exhibit similar SH intensities. We further measured the SH generation as a function of the linear polarization orientation, as depicted in Fig. S4d. As the input polarization rotating for 360 degrees, the SH generated in the perpendicular and parallel polarization states are alternatively enhanced as two dipolar-like patterns. This effect ensures the efficient generation of SH for all incident polarizations.

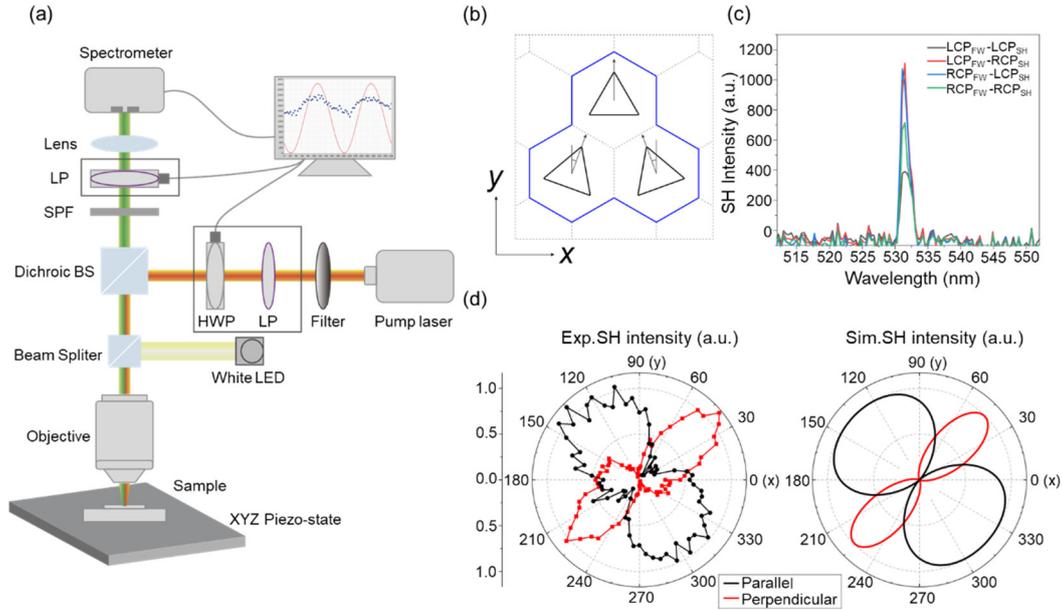

*Fig. S4. (a) Experimental setup for polarization-dependent SH. HWP, half-wave plate; SPF, short pass filter. (b) Schematic representation of the measured superlattice. (c) Polarization-dependent spectra of the SH from the PCSs. The pumping wavelength is ~1064 nm. Four combinations of circular polarization states of the FW and the SH wave were recorded. (d) Polar plot of the normalized SH intensity as a function of the incident polarization angle, with detected SH polarization perpendicular (red line) and parallel (black line) to the fundamental wave. The right panel shows the results from theoretical calculations.*

**Supplementary Section S5: Structural characterization**

The fabrication quality of the photonic crystal was characterized by optical microscopy and scanning electron microscopy (SEM, Zeiss Ultra Plus). Figure S5a shows an optical image of the PCSs with varying radii $R$ and rotation angle $\delta_B$. Figure S5b shows a zoomed-in SEM image depicting a portion of six fabricated photonic crystals with different angles $\delta_B = 0°$, $1°$, $2°$, $5°$, $10°$, and $20°$, respectively. The circumradius of the triangle hole is 145 nm. The red polygons indicate the supercell of the PCSs.

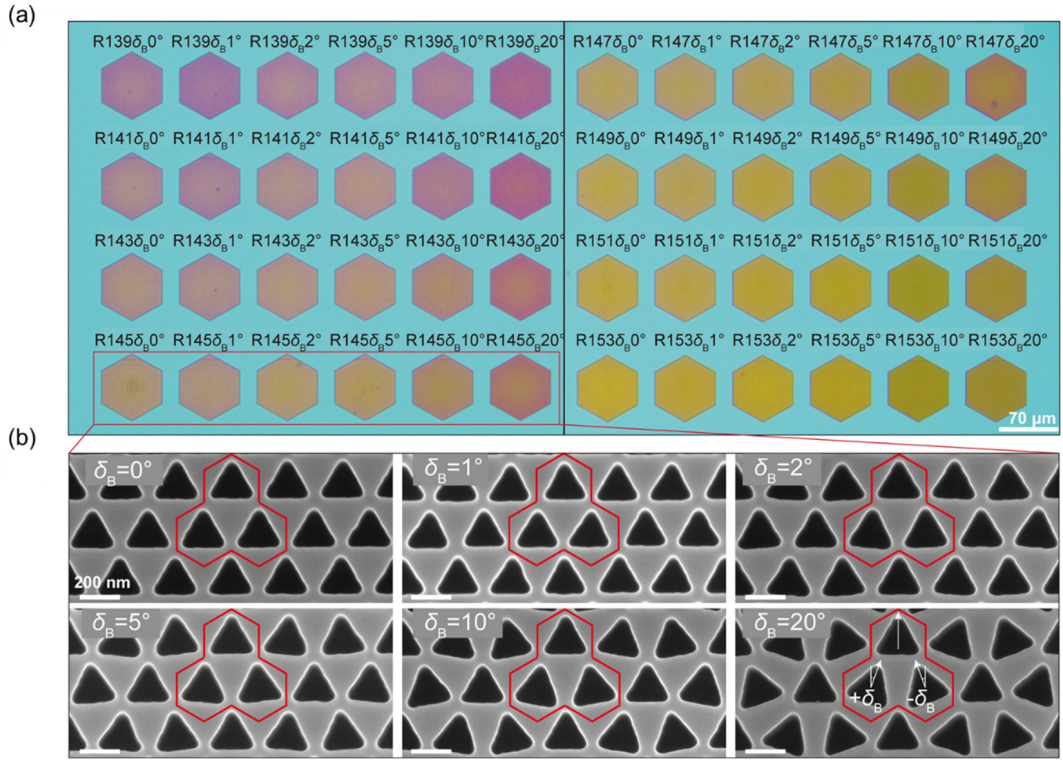

**Fig. S5**. *Structural characterization of some fabricated PCSs. (a) Optical images; (b) SEM images. Scale bars, 200 nm.*

**Supplementary Section S6: SH intensity as a function of exposure time**

Because the SH intensities generated from different elements are drastically varying across several orders of magnitudes, it is very challenging to correctly measure the SH signals from different PCS with the same pump power and detector exposure time. In order to precisely predict the enhancement of SH generation, we calibrated the response linearity of the detector as a function of exposure time. The results are shown in Fig. S6, indicating that the counts are linearly increasing with the exposure time. Therefore, the calibration in Fig. S6 allows us to adjust the exposure settings of the detector and laser power, to correctly obtain the SH enhancements from different PCSs spanning several orders of magnitudes.

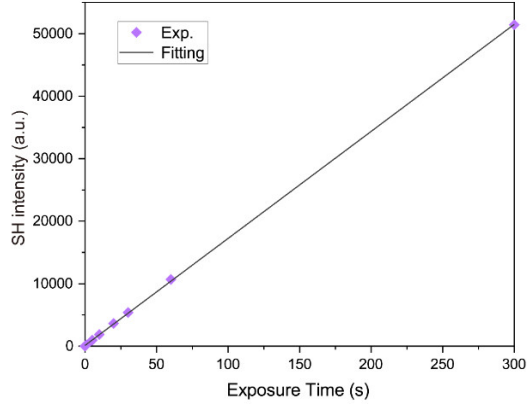

***Fig. S6***. *Calibration of the measured SH intensity as a function of the detector exposure time. The purpose of this measurement is to evaluate the SH enhancement from different samples, which can vary across several orders of magnitudes.*

**Supplementary Section S7: Symmetry properties of our supercell PCS**

To clarify the nearfield vortex dynamic of our supercell PCS, we performed symmetry analysis in the parameter space $(\delta_T, \delta_B)$. The rotation of $\delta_B$ and $\delta_T$ results in the supercell structure, which reduces the Brillion zone: the original valley states at K(K′) point are therefore folded onto the Γ point. In this two-dimensional parameter space $(\delta_T, \delta_B)$, a double mode degeneracy always remains at the Γ point, allowing us to define a pair of pseudospins in the circular polarization basis. As it shows in Fig. S7, the vortex dynamic and FM are strictly constrained by the symmetry properties of the rotation operations. The mirror-symmetry axes of the top triangle changes along with $\delta_T$, as it shows by the red lines of Fig. S7. However, the mirror-symmetry axes for the bottom triangles are forced to be $\sigma_1$, $\sigma_2$ and $\sigma_3$, as shown by the black dashed lines. These symmetry axes will not change with $\delta_B$ because of our setting of opposite rotations of the bottom triangles. Therefore, the symmetry for the top triangle and bottom triangles are mismatched, and the point group and space group for the supercell structure are $C_3$ and $p_3$, respectively. Only if $\delta_T = m\frac{\pi}{3}, m \in \mathbb{Z}$, the symmetry is matched, and the supercell obtains a higher symmetry ($C_{3v}^d$, $p31m$) [1]. Because of these constrains, we can only observe a merged vortex for $\delta_T = m\frac{\pi}{3}$. In this scenario, the free vortices are moving along the mirror-symmetry axes (the black dashed lines). The $\delta_B$ tunes the coupling strength between the triangles, which also controls the moving speed of free vortices to merge at a critical angle $\delta_B = \delta_{cri}$, resulting in frustrated mode.

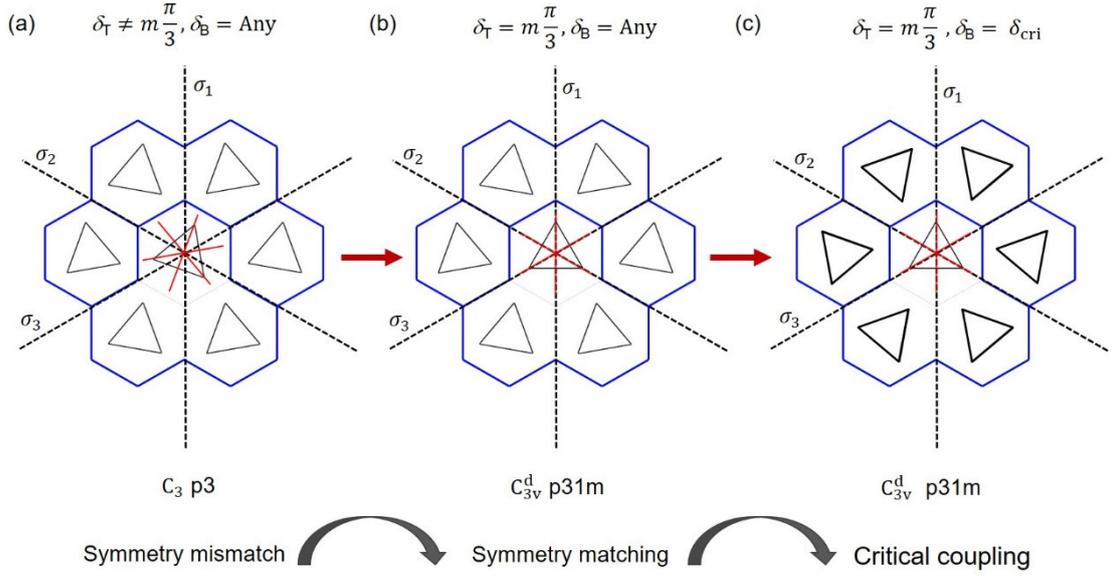

**Fig. S7.** *Symmetry analysis of the supercell PCS. (a) For $\delta_T \neq m\pi/3$, the mirror symmetries for the top triangle and bottom triangles are mismathced; (a) For $\delta_T = m\pi/3$, the mirror symmetries for the top triangle and bottom triangles arematchced, allowing for a FM emerging from a critical point, as indicated in (c).*

**Supplementary Section S8: Detailed properties of the supercell Bloch modes**

In Fig. S8. We show the nearfield properties of different Bloch modes ($\Gamma_{12}$) in an array of supercell structures. These Bloch modes are distinguished by four stages from their nearfield distributions: Valley, qBIC, FM and qValley. The Valley mode is exhibited as opposite vortices alternatively and equally assigned in the nearfield. The increase of $\delta_B$ causes the free vortices (dots) to move towards the pseudospins (the centers of triangles), as shown in Fig. S7 (a). Generally, q-BICs are small deviations from the valley modes, which can be explained by the small displacement of nearfield free vortices. FM corresponds to the vortex merging condition. For $\delta_B = 60°$, the vortices form a distribution which is very similar to that of a valley mode, and hence it is called a quasi - Valley mode (q-Valley). In Fig. S8 (b), we also show the in-plane Poynting vector $\boldsymbol{P} = Re\{\boldsymbol{E}^* \times \boldsymbol{H}\}/2$ and intensity |E| distributions. It can be seen that for the valley mode ($\delta_B = 0°$), the power flow forms small periodical vortices, the centers of which overlap with the free vortices. The intensity distribution of the Bloch modes has a unit-cell periodicity. As $\delta_B$ increases, the oppositely flowing Poynting vortices are merged to form a large vortex around the top triangles [Fig. S8(b), third]. Meanwhile, the spatial asymmetry of the supercell is maximized, as can be seen from the largest intensity contrast between the top triangles and bottom triangles. Interestingly, the q-Valley mode has a similar Poynting vortex formation to the valley mode, but with a supercell intensity periodicity.

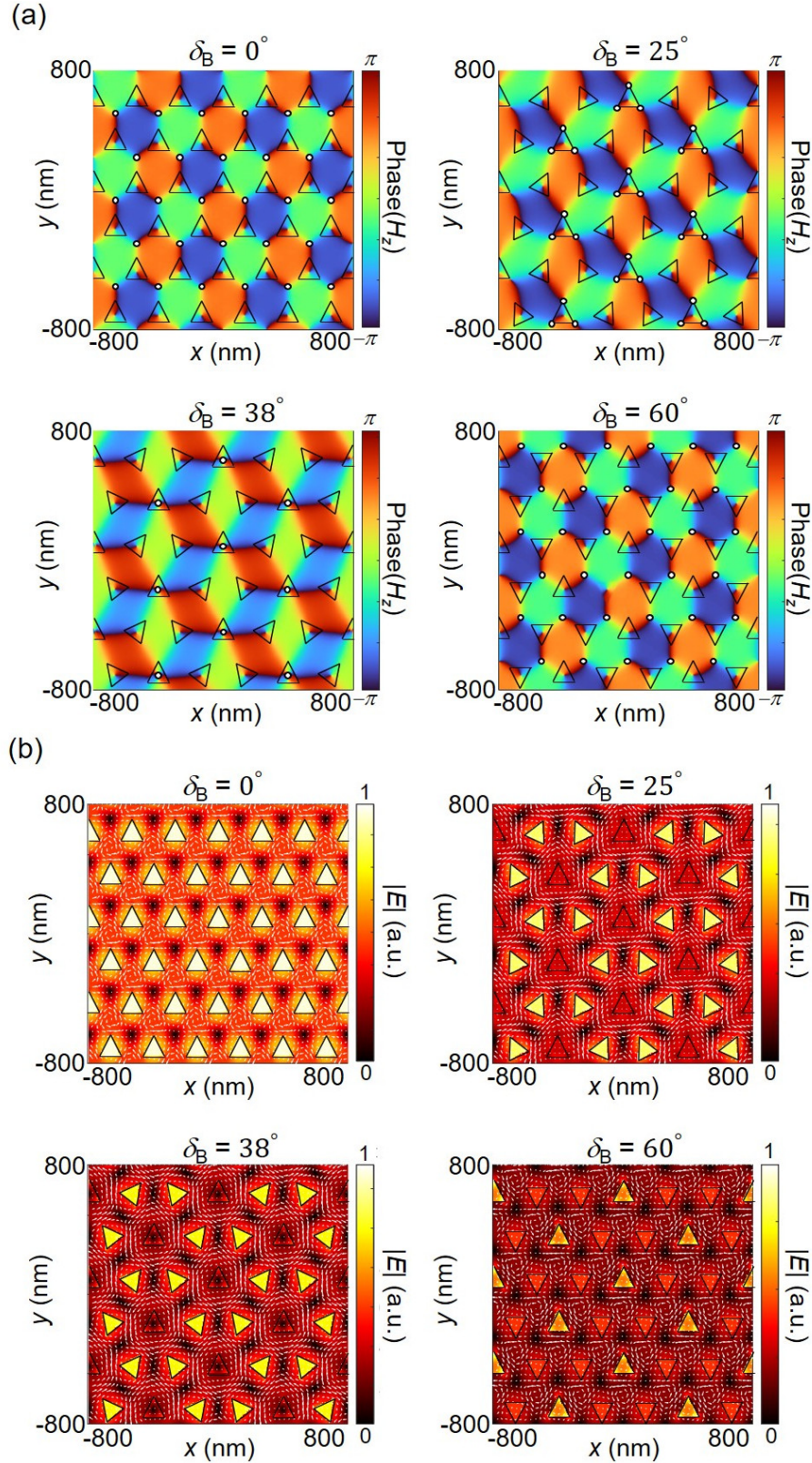

***Fig. S8.*** *Details for the nearfields of Bloch modes at different states. (a) vortex distributions; the white dots are free vortices; (b) Poynting vector and intensity distributions. The Poynting vectors are shown by the white arrows. In the simulations we use $\delta_T = 0$.*

**Supplementary Section S9: Temporal coupled mode theory for SHG**

To evaluate the SH enhancement from our supercell Bloch mode, we used the temporal coupled mode theory (TCMT) to consider the nonlinear resonate cavity with a fundamental frequency mode $\omega_1$, SH mode $\omega_2$, and their nonlinear interaction. Giving the fundamental constrains of TCMT ( energy conservation, time reversal symmetry and reciprocity) and weak nonlinear coupling, the coupling equations shall be written as [2,3]:

$$\frac{da_1}{dt} = (i\omega_1 - \gamma_1)a_1 - i\omega_1\beta_1 a_1^* a_2 + \sqrt{2\gamma_{1r}}s_{1+}$$

$$\frac{da_2}{dt} = (i\omega_2 - \gamma_2)a_2 - i\omega_1\beta_2 a_1^2$$

$$s_{1-} = -s_{1+} + \sqrt{2\gamma_{1r}}a_1$$

$$s_{2-} = \sqrt{2\gamma_{2r}}a_2.$$

Here, $a_{1,2}$ are the complex amplitudes of the fundament/SH modes, $s_{1,2\pm}$ are the time-dependent amplitudes for input/output ports, $\gamma_{1,2}$ are the losses, which can be divided into radiation loss $\gamma_r$ and non-radiation loss $\gamma_{nr}$ (such as absorption):

$$\gamma = \gamma_r + \gamma_{nr}$$

The quality facto is $Q = \omega/2\gamma$, therefore:

$$\frac{1}{Q} = \frac{1}{Q_r} + \frac{1}{Q_{nr}}$$

According to the perturbation theory, we can have the nonlinear coupling factor $\beta_1$

$$\beta_1 = \frac{1}{4} \frac{\int dr \varepsilon_0 \sum_{ijk} \chi_{ijk}^{(2)}(r) \left(E_{1i}^* E_{2j} E_{1k}^* + E_{1i}^* E_{1j}^* E_{2k}\right)}{(\int dr \varepsilon_0 \varepsilon_1(r) |E_1|^2) \sqrt{\int dr \varepsilon_0 \varepsilon_2(r) |E_2|^2}}$$

Because of the energy conservation during the nonlinear coupling, $\beta_2 = \beta_1^*/2$. For our silicon PCS, we consider an effective $\chi_{eff}^{(2)}$ emerging from a thin layer at the air-silicon interface, and obtain the nonlinear overlapping factor:

$$\beta = \frac{\iint dr E_2^* \cdot E_1^2}{\iint dr \varepsilon_1(r) |E_1|^2 \sqrt{\iint dr \varepsilon_2(r) |E_2|^2}}$$

Ignoring the down conversion process $\beta_1$, we have

$$\eta = \frac{P_{out}}{P_{in}^2} \propto |\beta|^2 \times \left[\frac{\gamma_1^2}{(\omega-\omega_1)^2+\gamma_1^2}\right]^2 \times \frac{\gamma_{1r}^2}{\gamma_1^4} \times \frac{\gamma_2^2}{(2\omega-\omega_2)^2+\gamma_2^2} \times \frac{\gamma_{2r}}{\gamma_2^2}$$

Combining above equations, we have

$$\eta \propto |\beta|^2 L_1^2(\omega-\omega_1) L_2(2\omega-\omega_2) \frac{Q_1^4 Q_2^2}{Q_{1r}^2 Q_{2r}}$$

Here $L_{1/2}$ are the Lorentz line functions.

In our PCS, the fundamental modes have large $Q$ factors and negligible absorption, so $Q_{1r} \approx Q_1$. Meanwhile, $Q_2$ is small due to absorption. Considering an approximation, $\omega \approx \omega_1 \approx \omega_2/2$, we have

$$\eta \propto Q_1^2 |\beta|^2 \qquad (1)$$

Equation (1) shows that in our PCS, the SH efficiency is determined by the $Q$-factor of fundamental mode and the nonlinear overlapping factor.

We used COMSOL to calculate the $Q_1$, $Q_2$, $|\beta|$ and $|\kappa_1|$ as a function of $\delta_B$, in order to compare our simulation results with the TCMT. Here, the quality factors are extracted by the eigen value solvers, and the $\beta$ factors are calculated by the integral between polarized plane waves and the excited nearfield modes. As it shows in Fig. S9(a), $|\beta|$ reaches a maximum around the FM, indicating that the nearfield vortex merging induces a maximum nonlinear coupling between the fundamental mode and SH mode on the PCS surface. On the other hand, $Q_1$ decreases along with $\delta_B$, and it reaches a minimal around the FM, a positive topological charge $q$. This FM is drastically different from qBIC, which is almost decoupled with free space optical modes from extremely high $Q$-factors. Fig. S9 (d) shows the $Q$-factors of the Bloch modes around $\omega_2 \approx 2\omega_1$. Unlike the fundamental modes, these modes are complex due to their small wavelengths. Roughly speaking, $Q_2$ can be considered as a constant (~100) as $\delta_B$ changes. Therefore, $Q_2$ does not affect the SH generation in our PCSs.

The abovementioned analysis revealed a different strategy to enhance SH in our PCSs: we use the rotation of structures to induce optical FM, which maximize the $Q_1\beta$ factor to obtain optimized SH from a large range of parameter change. This simple strategy can be used in more complex systems of different materials, within which by simply rotating the structures one can find an optimized solution to maximize $Q_1\beta$. Unlike the conventional approaches to obtain high-$Q$ mode from weak perturbation, such as qBIC, our FM is observed far away from a perturbation region, as the merged vortices are strongly deviated from their original locations (valley modes).

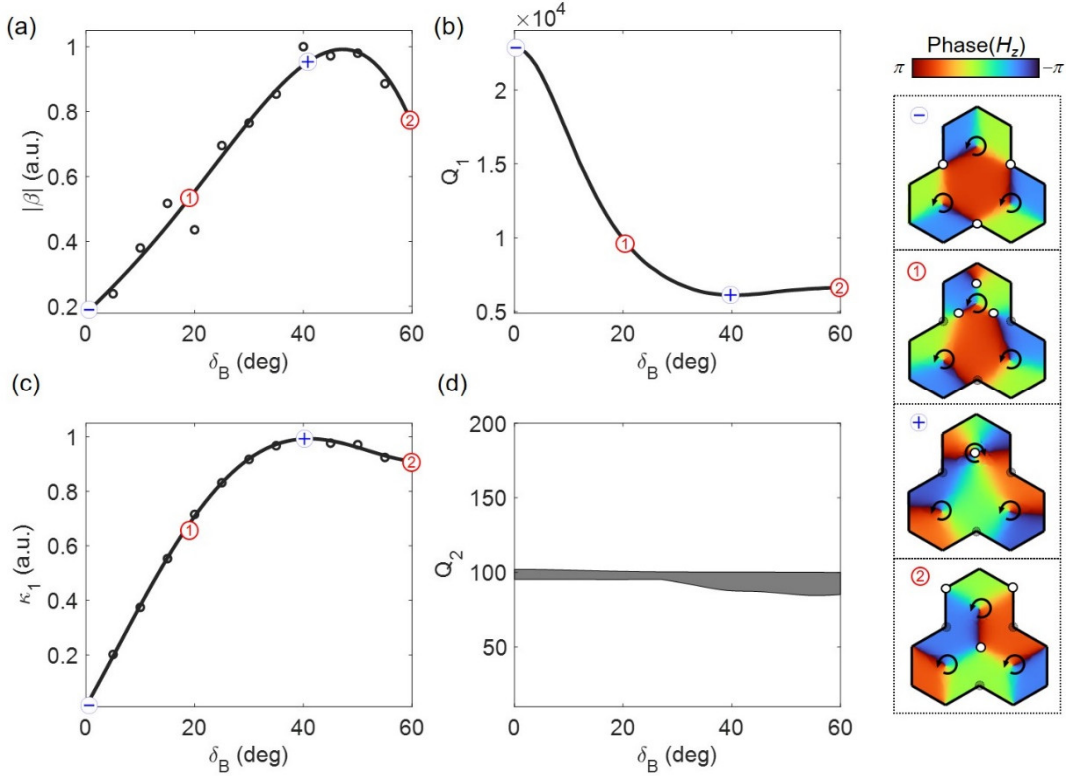

*Fig. S9.* *The simulated Q-factors for the fundamental mode and SH mode, nonlinear overlapping factor, and radiation coupling factor, as a function of $\delta_B$.*

**Supplementary Section S10: Simulated SH enhancement from different geometric parameters**

We use simulation to show that by varying the geometry parameters of our PCS, particularly $\delta_B$ and $R$, we can find optimized SH in a large parameter domain for $\delta_B \in (20°, 60°)$. This means that even from large fabrication inaccuracy, our structure can result in robust SH performance around the FM. Figure S10 represents the calculated SH from different combinations of $(\delta_B, R)$. These calculations are based on similar experimental parameters from our main text results, such as laser wavelength and SOI thickness. Generally, as $\delta_B$ increases, the SH also increases, agreeing with our experimental results. For different triangle sizes, the highest SH usually occurs for $\delta_B \in (20°, 60°)$.

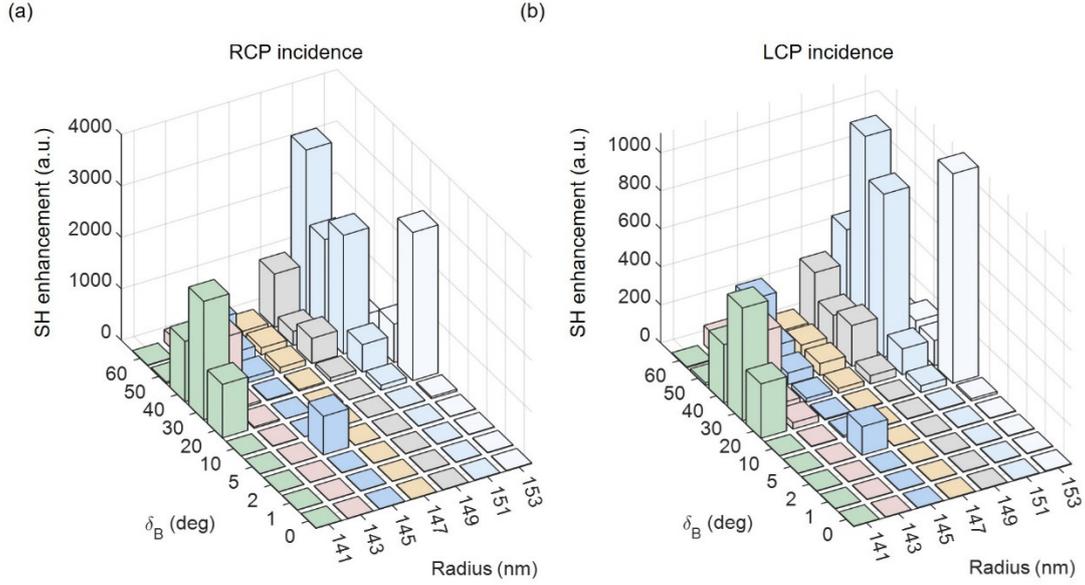

**Fig. S10.** *Simulated SH power as a function of R and $\delta_B$ under different circular polarization pump.*

**Supplementary Section S11: Comparing the properties of our supercell PCS with a $C_{6v}$-PCS**

We compare our structure with a different honeycomb PCS that induces supercell from size perturbation. Notably, our PCS is $C_{3v}$ symmetric, because of the Brillion folding from the K/K′ valley modes to the Γ-point modes. As it depicts in Fig. S11 (a), there is another possible approach to induce supercell by altering the size of circular holes in a honeycomb PCS: reducing the radius of the top hole by $-dr$ and increasing the radii of the bottom holes by $+0.5dr$ (Fig. S11). By this means, we constructed a $C_{6v}$-PCS, which also has a pair of Γ-point modes. Here, we list the main differences between such $C_{6v}$-PCS and our $C_{3v}$-PCS [FigS11 (d)]:

i. The two pairs of Γ-point modes in the $C_{6v}$-PCS cross as $dr/r_0$ changes, resulting in a topological phase transition. In contrast, as $\delta_B$ changes, the two pairs of Γ-point modes in the $C_{3v}$-PCS will never cross, remaining a large band gap.

ii. Considering the nearfield vortex distribution in the circular polarization basis, one pair of the $C_{6v}$-PCS is always FM (Fig. S11(b), red), while the other is always quasi-valley mode [Fig. S11(b), blue]. As for our $C_{3v}$-PCS, there is a vortex dynamic driven by $\delta_B$, and FM only occurs for a critical $\delta_B$. When $\delta_B > 60°$, FM transits from the lower $\Gamma_{12}$ band to higher $\Gamma_{34}$ band [Fig. S11(e)].

iii. The FMs in $C_{6v}$-PCS modes are always BIC modes (FigS11(c), red), with extremely

high Q factor. As for our $C_{3v}$-PCS, only the valley modes ($\delta_B = 0°$) have infinity $Q$ factors.

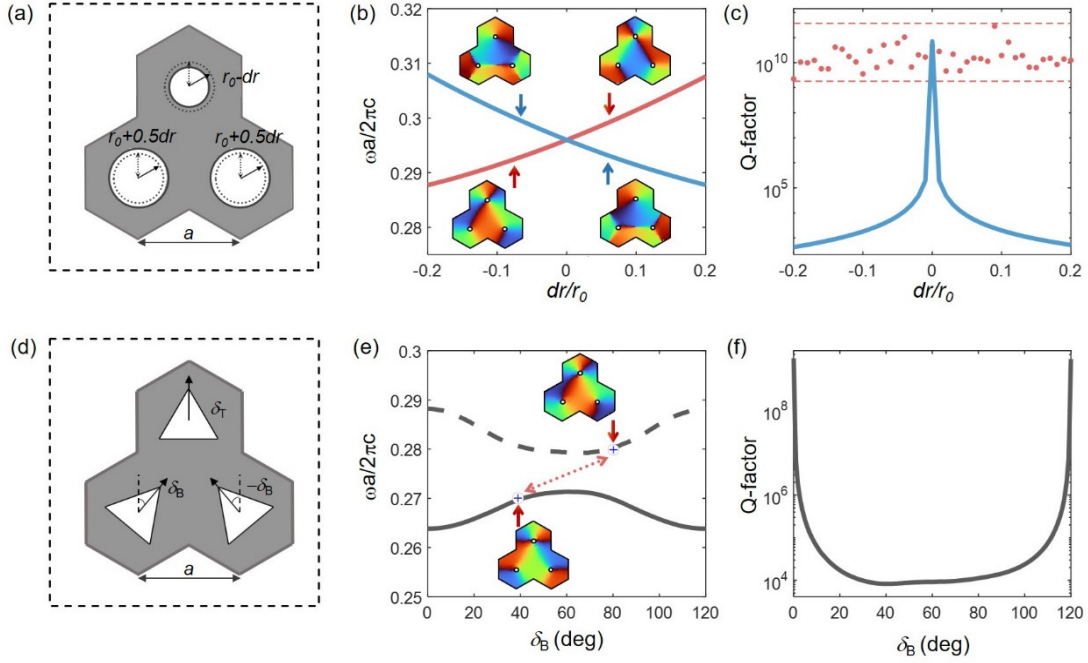

**Fig. S11**. Comparison between our $C_{3v}$-PCS with a $C_{6v}$-PCS. (a) A sketched supercell for $C_{6v}$-PCS; (b) The evolution of two pairs of Γ-point modes as a function of $dr/r_0$. The insets show the nearfield vortex distributions for the two types of modes; (c) The $Q$-factors of the two Γ-point modes as a function of $dr/r_0$. (d) A sketched supercell for our $C_{3v}$-PCS; (e) The evolution of a pair of Γ-point modes as a function of $\delta_B$. The insets show the nearfield vortex distributions for the two modes; (f) The $Q$-factors of one Γ-point mode as a function of $\delta_B$.

**Supplementary Section S12: Momentum space polarization vortices for $\Gamma_{1,2}$ Bands**

Alongside the nearfield vortices for Γ-point modes, we also show the momentum space polarization textures of these modes around the Γ-point, as depicted in Fig. S12. It can be seen that the center of momentum space is always a pair of orthogonal linear polarizations, which can be decomposed into a pair of circular polarization states to observe nearfield vortices. For other modes away from the Γ-point, this degeneracy is removed. The polarization distribution of the momentum space is restricted by the mirror-symmetry that was mentioned in Fig. S7, which are divided by the L-lines.

For $\delta_B = 5°$ and $\delta_B = 60°$, the momentum space distribution is closer to that from a valley mode and qValley mode, which allows for circular polarizations to emerge in the momentum space, in corresponding to $C$ points.

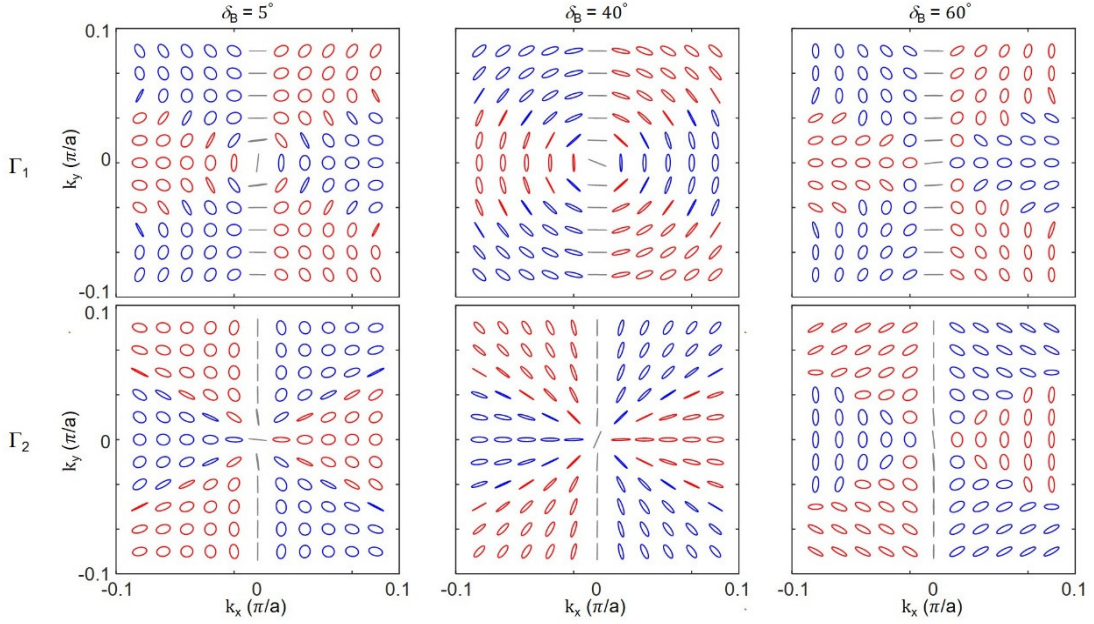

**Fig. S12.** *Momentum-space polarization distributions of the $\Gamma_{1,2}$ bands for several different $\delta_B$.*